\documentclass[a4paper]{ESASPCS13Style}
\usepackage{epsfig}

\begin{document}

\title{Newly discovered active binaries in the RasTyc sample of stellar X-ray sources}

\author{A. Frasca\inst{1}, P. Guillout\inst{2}, E. Marilli\inst{1}, R. Freire Ferrero\inst{2}, K. Biazzo\inst{3}}
  \institute{INAF - Osservatorio Astrofisico di Catania, via S. Sofia, 78, 95123 Catania, Italy
\and Observatoire Astronomique de Strasbourg- ULP, 11 rue de l'Universit\'e, 67000 Strasbourg, France 
\and Dipartimento di Fisica e Astronomia, Universit\`a di Catania, via S. Sofia, 78, 95123 Catania, Italy}

\maketitle 

\begin{abstract}

We present preliminary results of follow-up optical observations, both photometric and spectroscopic, of 
stellar X-ray sources, selected from the cross-correlation of ROSAT All-Sky Survey (RASS) and TYCHO catalogues.
Spectra were acquired with the E{\sc lodie} spectrograph at the 193-cm telescope of the Haute Provence 
Observatory (OHP) and with the REOSC echelle spectrograph at the 91-cm telescope of the Catania Astrophysical 
Observatory (OAC), while $UBV$ photometry was made at OAC with the same telescope. 
In this work, we report on the discovery of six late-type binaries, for which we have obtained good radial velocity 
curves and solved for their orbits. Thanks to the OHP and OAC spectra, we have also made a 
spectral classification of single-lined binaries
and we could give first estimates of the spectral types of the double-lined binaries. 
Filled-in or pure emission H$\alpha$ profiles, indicative of moderate or high level of chromospheric activity, 
have been observed.  We have also detected, in near all the 
systems, a photometric modulation ascribable to photospheric surface inhomogeneities which is correlated 
with the orbital period, suggesting a synchronization between rotational and orbital periods.
For some systems has been also detected a variation of H$\alpha$ line intensity, with a possible 
phase-dependent behavior. 

\keywords{Stars: binaries: spectroscopic  -- Stars: fundamental parameters -- Stars: X-ray -- Stars: activity}	
\end{abstract}

\section{Introduction}
\label{sec:Intro}

The cross-correlation between the ROSAT All-Sky Survey ($\simeq$ 150\,000 sources) and the  TYCHO mission 
($\simeq$ 1\,000\,000 stars) catalogues has selected about 14\,000 stellar X-ray sources (RasTyc sample, 
\cite{Guillout99}). Although most of these soft X-ray sources are expected to be the youngest stars in the 
solar neighborhood, neither the contamination by older RS~CVn systems nor the fraction of BY~Dra binaries are 
actually known. This information is, however, of fundamental importance for studying the recent local star formation 
history and, for instance, for putting constrains on the scale height of the spatial distribution of nearby young 
stars around the galactic plane. 
We thus started a  spectroscopic 
observation campaign aimed at a deep characterisation of a representative sub-sample of the RasTyc 
population. In addition to derive chromospheric activity levels (from H$\alpha$ emission) and rotational 
velocities (from Doppler broadening), high resolution spectroscopic observations allow to infer ages 
(by means of Lithium abundance) and to single out spectroscopic and active binaries. 
In this work we present some preliminary results of follow-up observations, both photometric and spectroscopic, 
of some RasTyc stars performed with the 193-cm telescope of OHP and the 91-cm telescope of the Catania 
Astrophysical Observatory (OAC). 

In particular, we analyse six new late-type binaries, for which we have obtained good radial velocity 
curves and orbital solutions. An accurate spectral classification for the single-lined binaries has been 
also performed and the projected rotational velocity $v\sin i$ has been measured for all stars. 
The chromospheric activity level and the lithium content have been also investigated using as 
diagnostics the H$\alpha$ emission and the Li{\sc i}\,$\lambda\,6708$ line, respectively.

\begin{table*}[bht]
  \caption{Properties of the systems}
  \label{tab:param}
  \begin{center}
    \leavevmode
    \footnotesize
    \begin{tabular}[h]{llcccccccr}
      \hline \\[-5pt]
RasTyc  &   Name  &   P$_{\rm orb}$  & 	$\gamma$  &  $k$ (P/S)  &  $M\sin^3i$  &  $v\sin i$ (P/S) &  Sp. Type  &  $B-V$  & 
W$_{\rm LiI}$ \\
        &         &   (days)         &    (km\,s$^{-1}$)  &  (km\,s$^{-1}$)   &  $M_{\odot}$  &    (km\,s$^{-1}$)      &            &         &
(m\AA)   \\[+5pt]
      \hline \\[-5pt]
193137  &  HD~183957	 & 7.954  &   $-$29.0  &   57.5/63.1   & 0.758/0.691  &   4.0/4.4     &     K0-1V/K1-2V  & 0.84  &    $< 10$ \\
215940  &  OT~Peg        & 1.748  &   $-$27.0  &   16.6/23.2   & 0.007/0.005  &   9.2/9.4     &     K0V/K3-5V	 & 0.79  &    50     \\
221428  &  BD+33\,4462   & 10.12  &   $-$20.9  &   59.2/60.4   & 0.905/0.887  &   16.1/32.6   &     G2 + K	 & 0.70  &    15:    \\
040542  &  DF~Cam        & 12.60  &   $-$19.5  &   22.8        & SB1	      &   35	      &     K2III        & 1.14  &    ---    \\
072133  &  V340~Gem	 & 36.20  &   +37.0    &   42.1        & SB1	      &   40	      &     G8III        & 0.83  &    70     \\
102623  &  BD+38\,2140   & 15.47  &   +47.4    &   31.3        & SB1	      &   11.5        &     K1IV         & 1.03  &    40     \\
      \hline \\
       \end{tabular}\\[-4pt]
  \end{center}
      \end{table*}

\section{Observations and reduction}
\label{sec:Obs}

\subsection{Spectroscopy}

  Spectroscopic observations have been obtained 
  at the {\it Observatoire de Haute Provence} 
  (OHP) and at the {\it M.G. Fracastoro} station (Mt. Etna, 1750 m a.s.l.) of Catania Astrophysical 
  Observatory (OAC).

 At OHP we observed in 2000 and 2001 with the E{\sc lodie} echelle spectrograph connected to the 193-cm 
 telescope.  The 67 orders recorded by the CCD detector cover the 3906-6818~\AA\ 
 wavelength range with a resolving power of about
 42\,000 (\cite{Bar96}).
 The E{\sc lodie} spectra were automatically reduced on-line during the observations and the 
 cross-correlation with a reference mask was produced as well.
 
  The observations carried out at Catania Observatory have been performed in 2001
  and 2002 with the REOSC echelle spectrograph at the 91-cm telescope. 
The spectrograph is fed by the telescope through an optical fiber
(UV - NIR, $200\,\mu m$ core diameter) and is placed in a stable position
in the room below the dome level. 
Spectra were recorded on a CCD camera equipped with a thinned back-illuminated
SITe CCD of 1024$\times$1024 pixels (size 24$\times$24 $\mu$m). The \'echelle crossed 
configuration yields a resolution of about 14\,000, as deduced from the FWHM 
of the lines of the Th-Ar calibration lamp. The observations have been made in the red 
region. The detector allows us to record five orders in each frame, spanning from about 5860
to 6700~\AA.

The OAC data reduction was performed by using the {\sc echelle} task of IRAF\footnote{IRAF is 
    distributed by the National Optical Astronomy Observatories,
    which are operated by the Association of Universities for Research
    in Astronomy, Inc., under cooperative agreement with the National
    Science Foundation.} 
package following the standard steps: background subtraction, division by a flat
 field spectrum given by a halogen lamp, wavelength calibration using the 
emission lines of a Th-Ar lamp, and normalization to the continuum through a polynomial fit.

\subsection{Photometry}
The photometric observations have been carried out in 2001 and 2002 in the standard $UBV$ 
system also with the 91-cm telescope of OAC and a photon-counting refrigerated photometer 
equipped with an EMI 9789QA photomultiplier, cooled to $-15\degr$C. The dark noise of the detector, 
operated at this temperature, is about $1$ photon/sec.

For each field of the RasTyc sources, we have chosen two or three stars with known $UVB$ magnitudes 
to be used as local standards for the determination of the photometric instrumental ``zero points". 
 Additionally, several standard stars, selected from the list of Landolt (\cite{Lan92}), 
were also observed during the run in order to determine the transformation coefficients 
to the Johnson standard system. 

A typical observation consisted of several integration cycles (from 1 to 3, depending on the star brightness) 
 of 10, 5, 5 seconds, in the $U$, $B$ and $V$ filter, respectively.
A 21$\arcsec$ diaphragm was used. The data were reduced by means of the photometric data
reduction package PHOT designed for photoelectric photometry of Catania Observatory 
(\cite{LoPr93}).
Seasonal mean extinction coefficient for Serra La Nave Observatory were adopted for 
the atmospheric extinction correction.

\section{Results}

\subsection{Radial velocity and photometry}
\label{sec:RV}

The radial velocity (RV) measurements for the E{\sc lodie} data have been performed onto
the cross-correlation functions (CCFs) produced on-line during the data acquisition.

Radial velocities for OAC spectra were obtained by cross-correlation of
each echelle spectral order of the RasTyc spectra  with that of bright radial velocity 
standard stars. 
 For this purpose the IRAF task {\sc fxcor}, that computes RVs by means of
 the cross-correlation technique, was used.

  The wavelength ranges for the cross-correlation 
were selected to exclude the H$\alpha$ and Na\,{\sc I} D$_2$ lines, which are contaminated
by chromospheric emission and have very broad wings. The spectral
regions heavily affected by telluric lines (e.g. the  O$_2$ lines in the 
$\lambda~6276-\lambda~6315$ region) were also excluded.

\begin{figure*}[ht]
  \begin{center}
    \epsfig{file=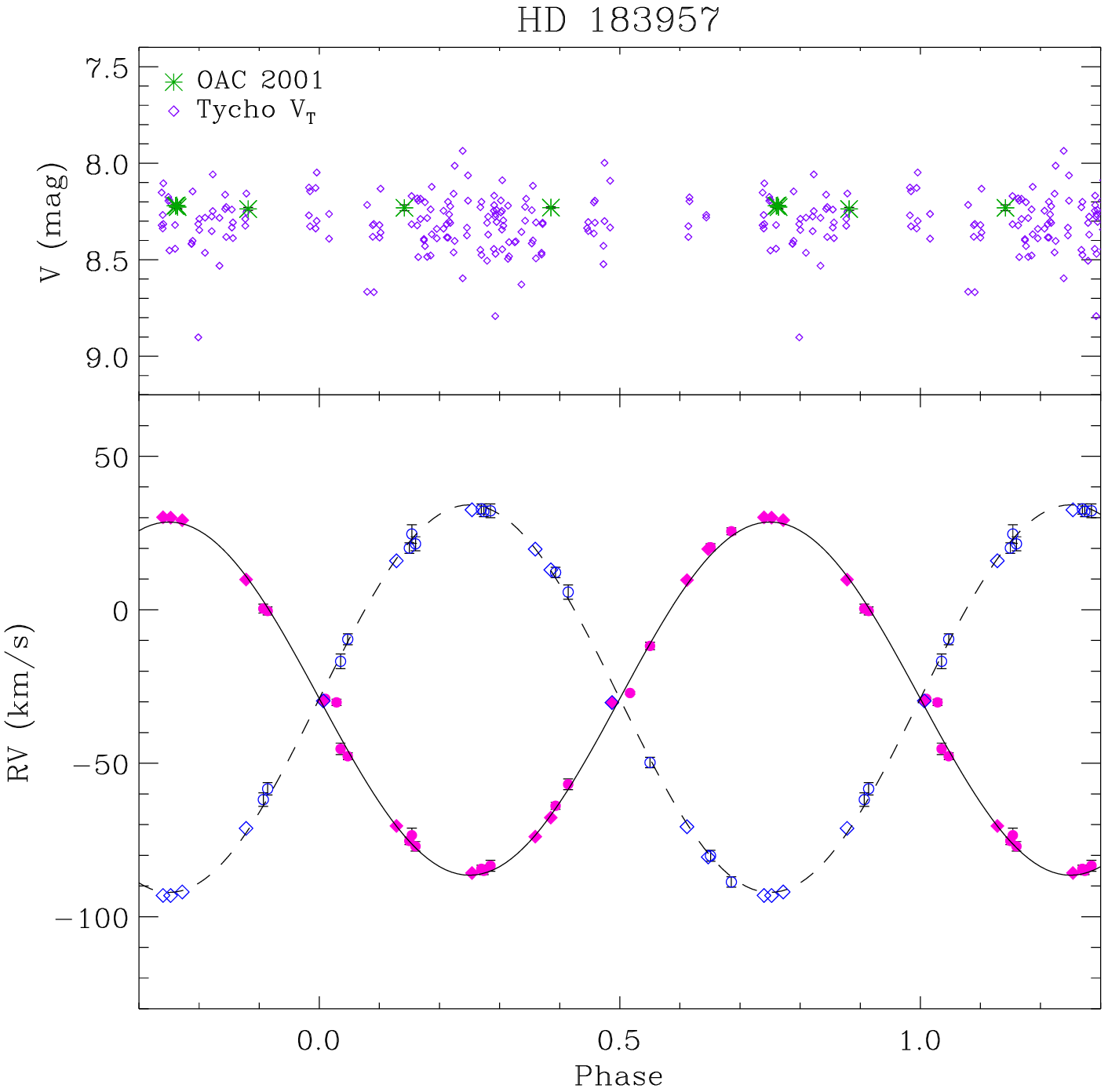,width=5.5cm}
  \epsfig{file=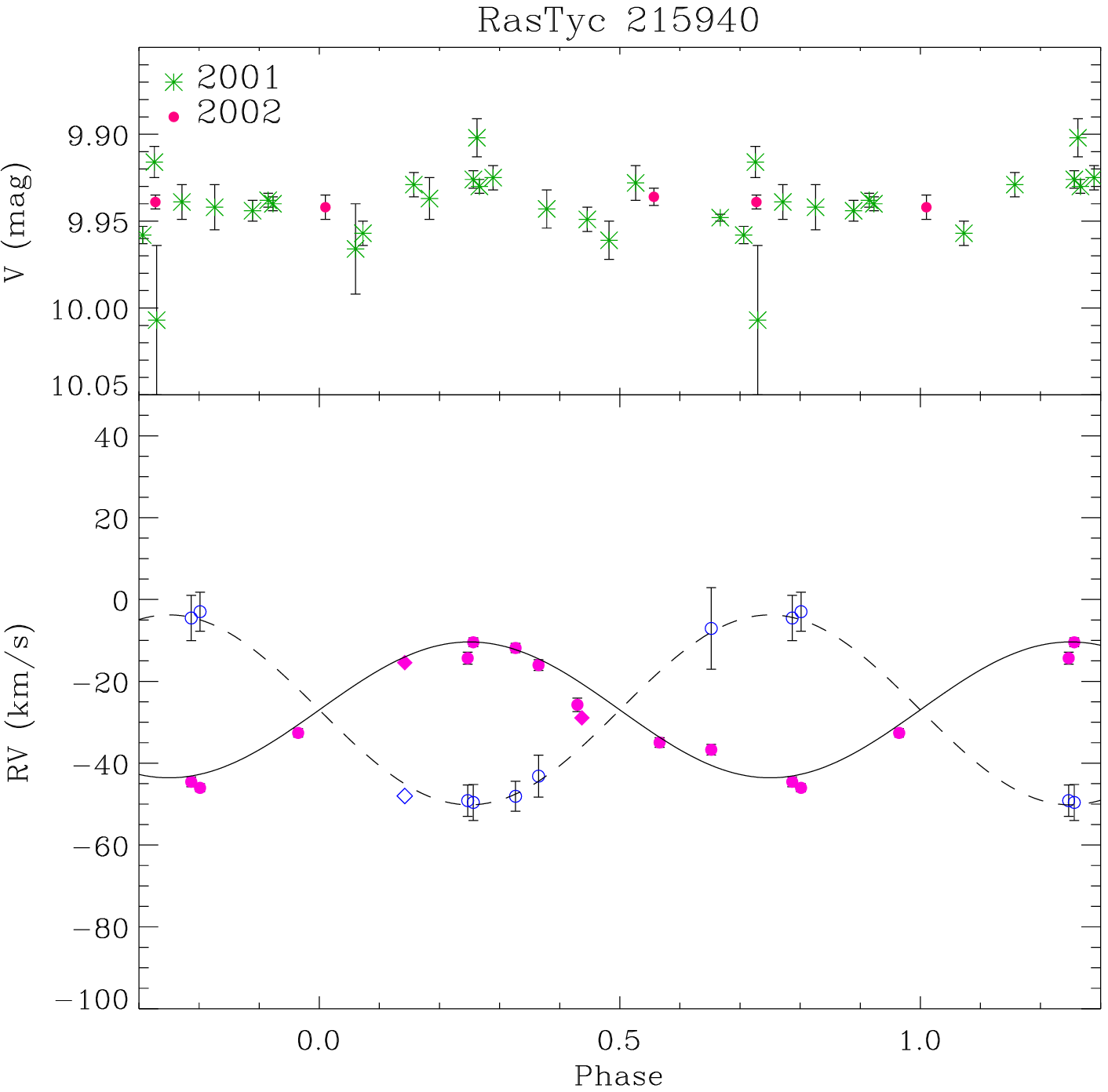,width=5.5cm}
  \epsfig{file=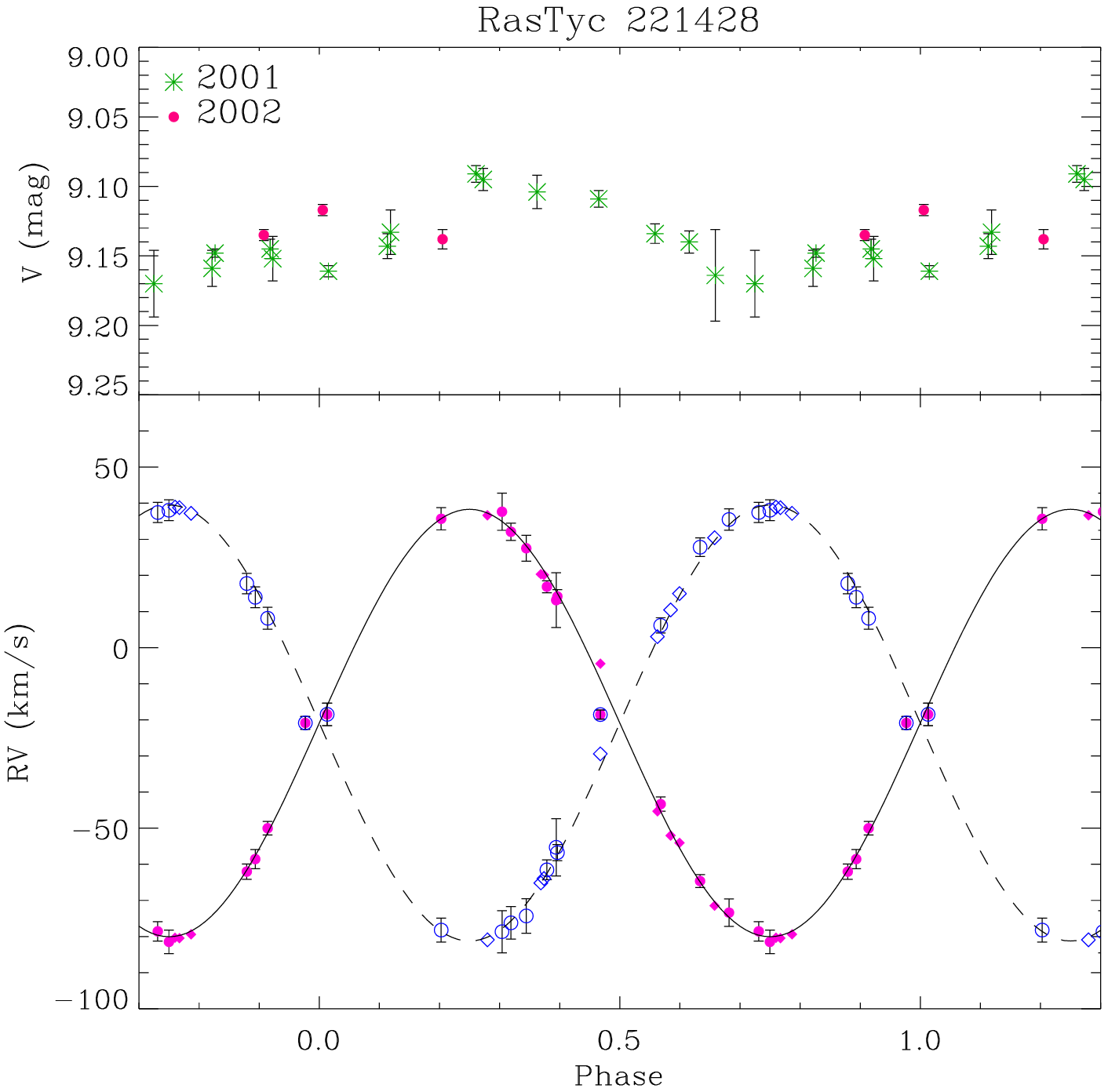,width=5.5cm}
  \epsfig{file=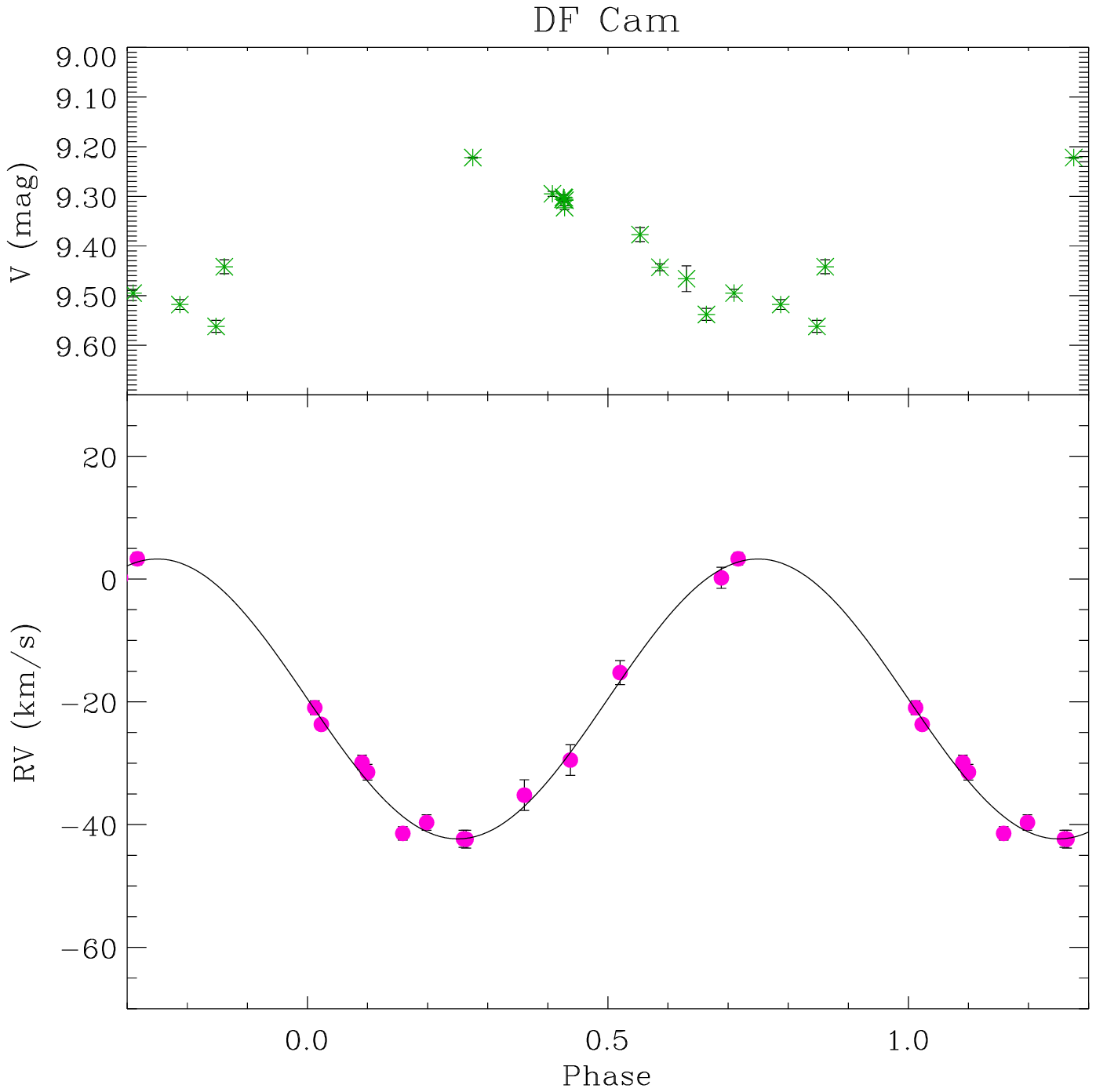,width=5.5cm}
  \epsfig{file=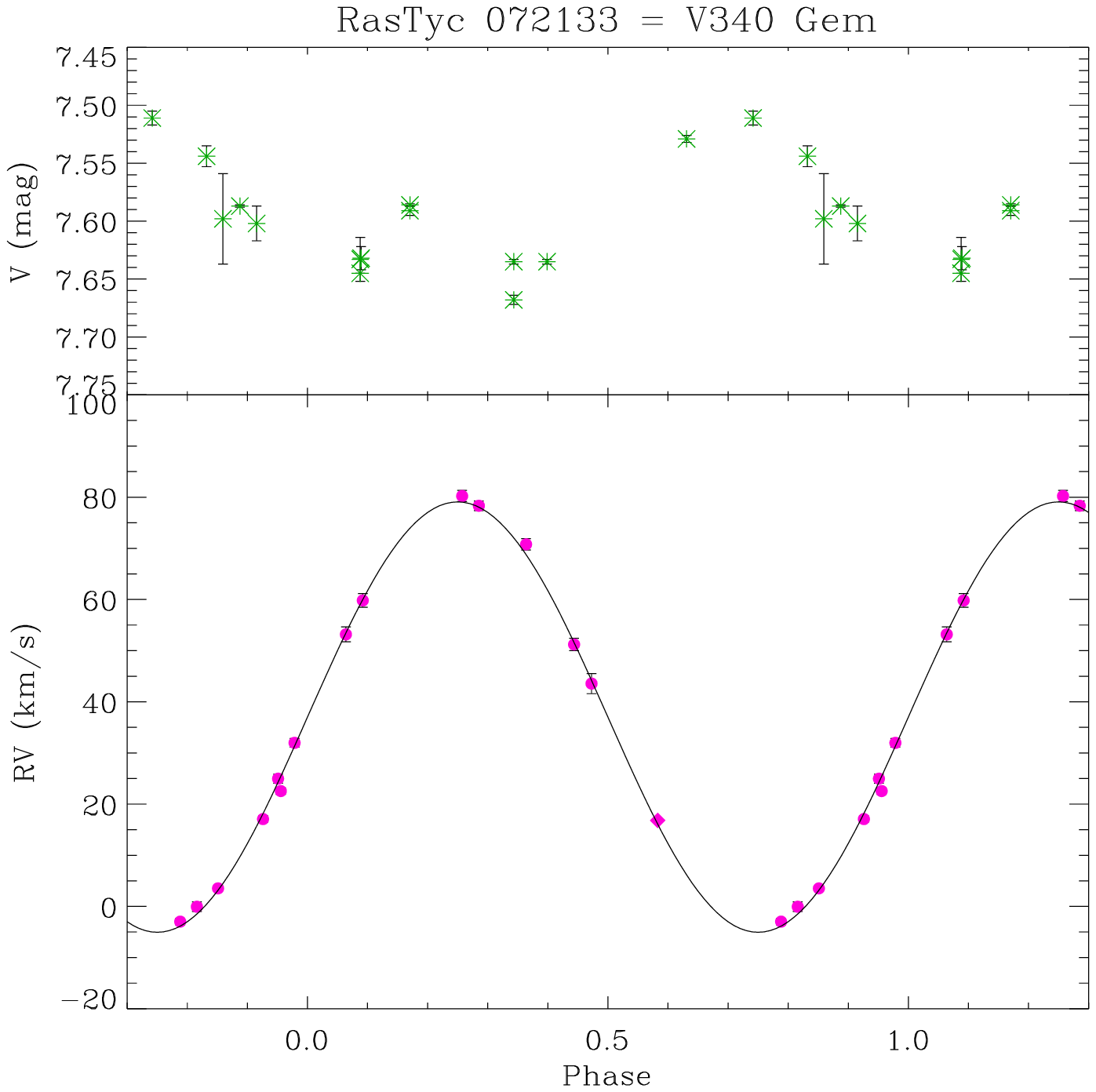,width=5.5cm}
  \epsfig{file=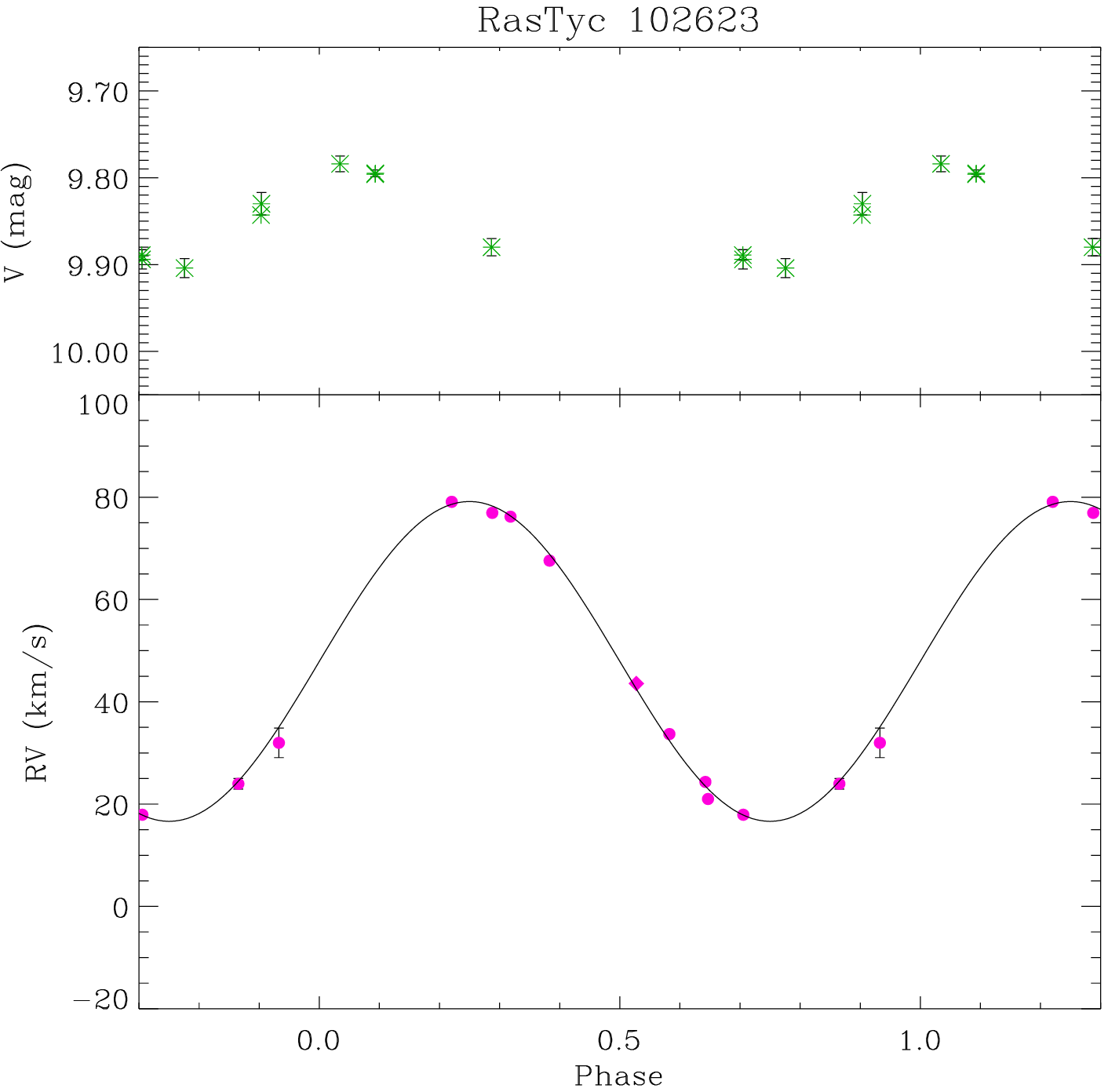,width=5.5cm}
  \end{center}
\caption{Radial velocity curves (dots= OAC data, diamonds= E{\sc lodie} data) of the six new RasTyc binaries. For the three SB2
systems we used filled and open symbols for the primary (more massive) and secondary components, respectively.
 The circular solutions are also represented with solid and dashed lines for the primary 
and secondary components, respectively. The contemporaneous V  photometry is displayed, as a function of 
the orbital phase, on the top panel of each box. }
\label{fig:RV}
\end{figure*}

  The observed RV curves  are displayed in
Fig.~\ref{fig:RV}, where, for SB2 systems, we used dots for the RVs of primary (more massive) components 
and open circles for those secondary (less massive) ones.
We initially searched for eccentric orbits  and found in any case very low eccentricity values (e.g. $e=0.010$
for HD~183957, $e=0.030$ for 221428). Thus, following the precepts of \cite*{Lucy71},
we adopted $e=0$.
 The circular solutions are also represented in Fig.~\ref{fig:RV} with solid and dashed lines for the primary 
and secondary components, respectively.

The orbital parameters of the systems, orbital period ($P_{\rm orb}$), barycentric velocity ($\gamma$), 
RV semi-amplitudes ($k$) and masses ($M\sin^3i$), 
are listed in Table~\ref{tab:param}, where P and S  refer to the primary and 
secondary components of the SB2 systems, respectively.

With the only exception of HD~183957, for which any modulation is visible neither in OAC data nor in TYCHO 
$V_{\rm T}$ magnitudes, all sources show a  photometric modulation well correlated with the orbital period,
indicating a high degree of synchronization. The low amplitude of the light 
curve of 215940 and the very low values of $M\sin^3i$ imply a very low inclination of orbital/rotational axis.

\subsection{Spectral type and $v\sin i$ determination}
\label{sec:Spty}
For SB1 systems observed with E{\sc lodie} we have determined effective temperatures and gravity (i.e. spectral 
classification) by means of the TGMET code, available at OHP (\cite{Katz98}). We have also used ROTFIT, a code 
written by one of us (\cite{Frasca03}) in IDL (Interactive Data Language, RSI), which simultaneously find 
the spectral type and the $v\sin i$ of the target by searching, into a library of standard star spectra, for 
the standard spectrum which gives the best match of the target one, after the rotational broadening.
As standard star library, we used a sub-sample of the stars 
of the TGMET list whose spectra were retrieved from the E{\sc lodie} Archive (\cite{Prugniel01}). 
The ROTFIT code was also applied to the OAC spectra, using standard star spectra acquired with the same instrument.
This was especially advantageous for DF~Cam, for which we have no E{\sc lodie} spectrum.

\begin{figure}[ht]
  \begin{center}
    \epsfig{file=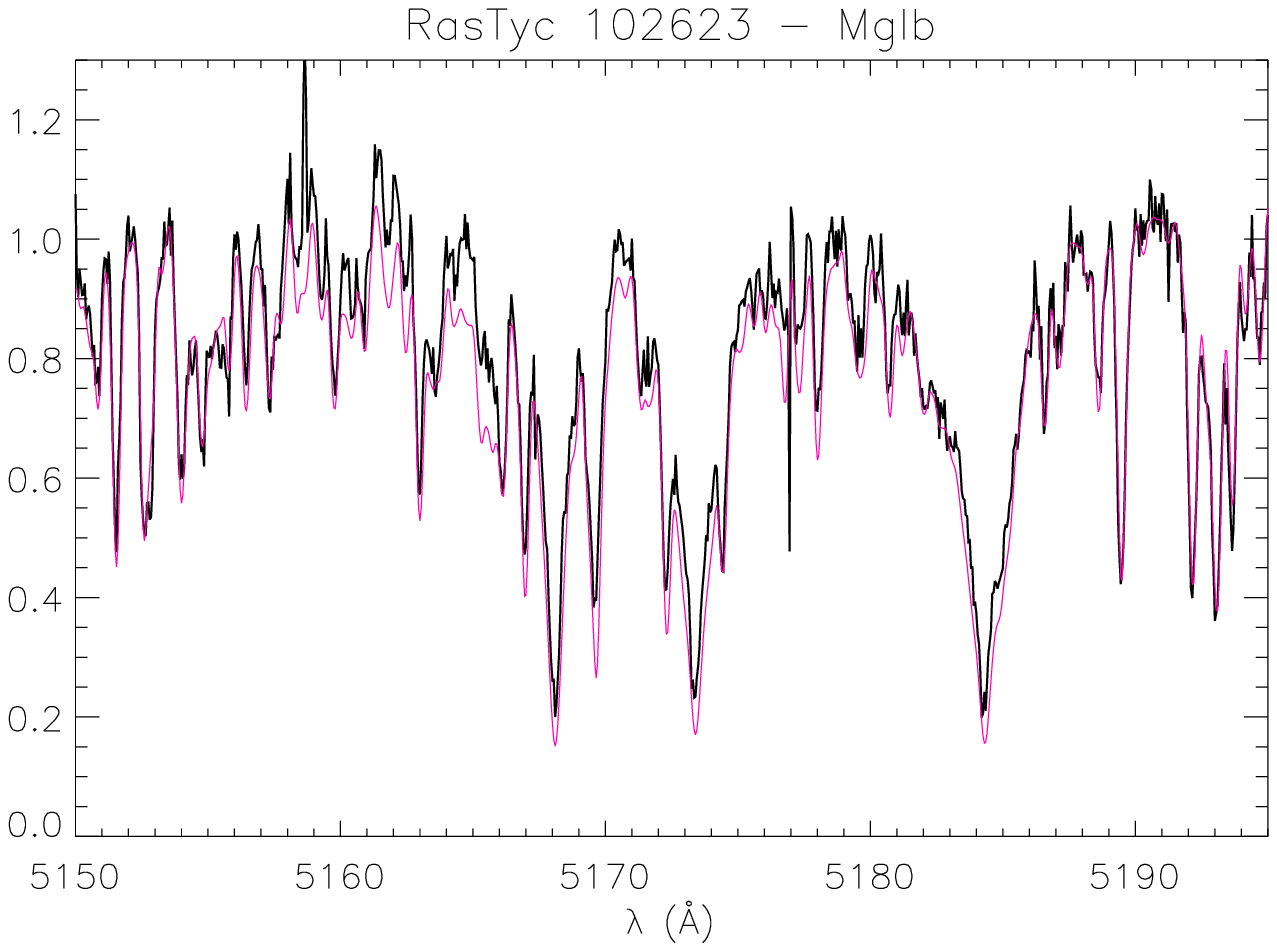,width=6.5cm,height=3.5cm}
  \epsfig{file=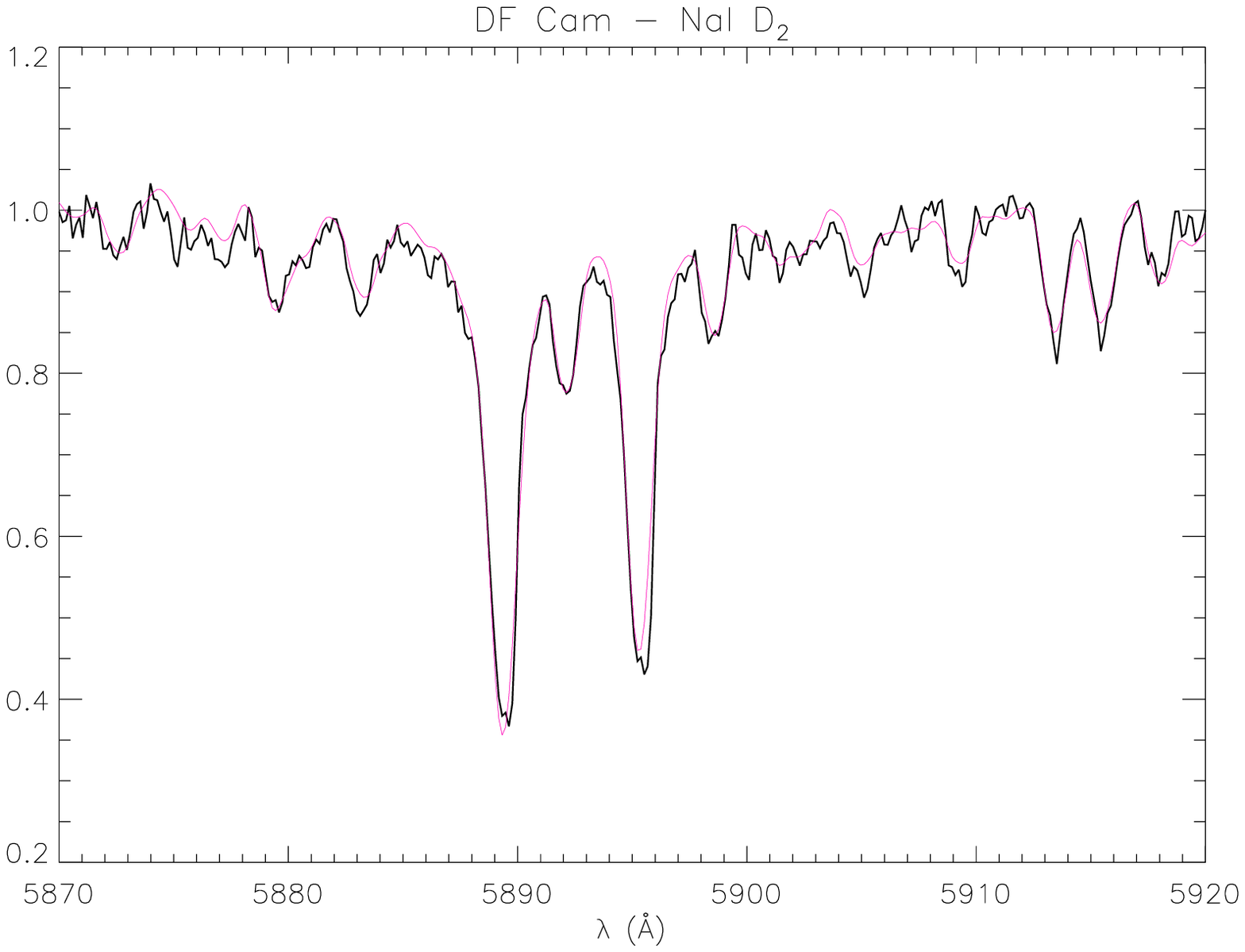,width=6.5cm,height=3.5cm}
  \end{center}
\caption{(Top panel) Observed E{\sc lodie} spectrum of RasTyc 102623 in the Mg{\sc i}b region (thick line) with the
superimposed spectrum (thin line) of the standard star which gives the best match ($\gamma$~Cep, K1\,IV) broadened at the $v\sin i$
of the target. 
(Bottom panel) OAC spectrum of DF~Cam in the Na{\sc i}\,D$_2$ region (thick line) with the superimposed spectrum 
(thin line) of the standard star $\alpha$~Ari (K2\,III) broadened at 35 km\,s$^{-1}$. }
\label{fig:synth}
\end{figure}

For SB2 systems we made a preliminary classification on the basis of a visual inspection of E{\sc lodie} and OAC 
spectra. However, we are developing a code for spectral type determination in double-lined binaries which
will allow us to improve the spectral classification. 
We found at least two binaries composed by main sequence stars, while the remaining systems contain an evolved
(giant or sub-giant) star.  

Measurements of $v\sin i$ were also made using the E{\sc lodie} CCFs and the calibration 
relation between CCF width and $v\sin i$ proposed by \cite*{Queloz98}. The lower rotation rate ($v\sin i \simeq$\,4
km\,s$^{-1}$) has been detected for both components of HD~183957, which display also the lowest H$\alpha$ 
activity among the six sources.

\subsection{H$\alpha$ emission and Lithium content}
\label{sec:Halpha}

The H$\alpha$ line is an important indicator of chromospheric activity.
Only the very active stars show always H$\alpha$ emission above the continuum, while
in less active stars only a filled-in
absorption line is observed. The detection of the chromospheric emission contribution 
filling in the line core is hampered in double-lined systems in which both spectra are
 simultaneously seen and shifted at different wavelengths, according to the orbital phase.
Therefore a comparison with an ``inactive'' template built up with two stellar
spectra that mimic the two components of the system in absence of activity is needed to emphasize
 the H$\alpha$ chromospheric emission. 

The inactive templates have been built up with rotationally broadened E{\sc lodie} archive spectra 
(HD~10476, K1\,V for both components of HD~183957; $\gamma$~Cep, K1\,IV for 102623; 
$\delta$~Boo, G8\,III for 072133; HD~17382, K1\,V for 215940) or with OAC spectra
of $\alpha$~Ari (K2\,III), for DF~Cam, acquired during the observing campaigns.

\begin{figure}[ht]
  \begin{center}
    \epsfig{file=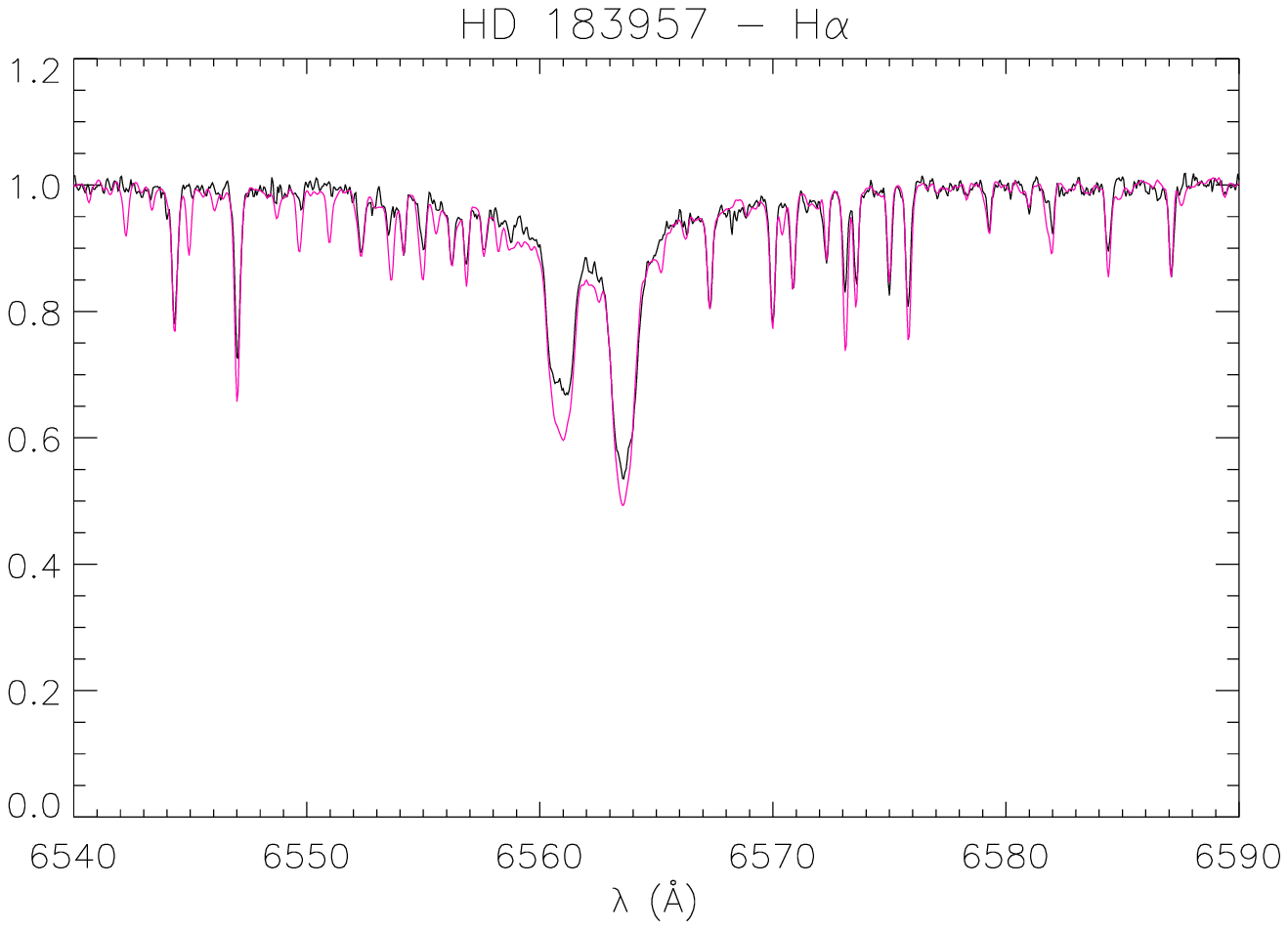,width=6.5cm,height=3.5cm}
  \epsfig{file=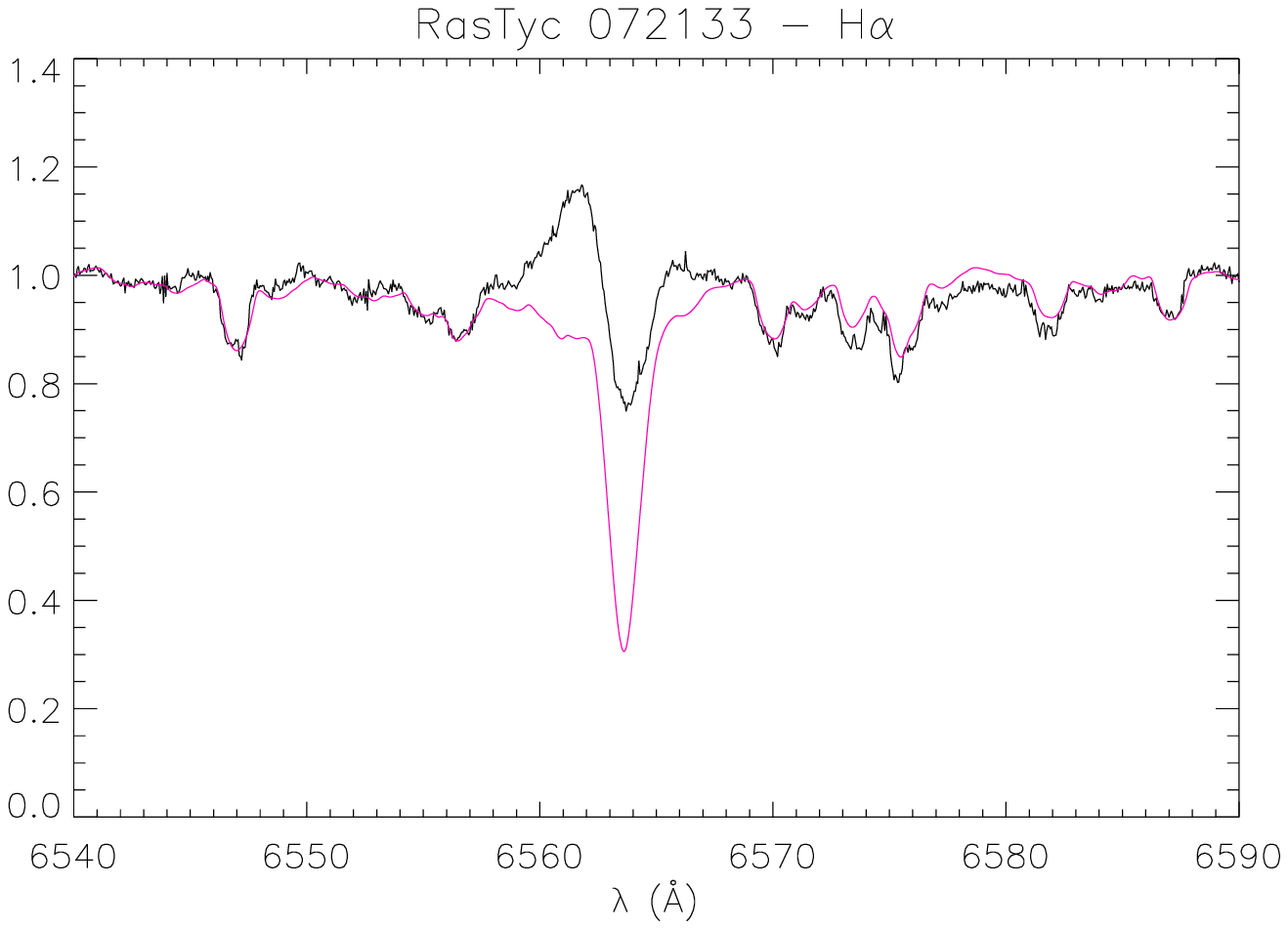,width=6.5cm,height=3.5cm}
  \end{center}
\caption{E{\sc lodie} spectra of HD~183957 and  RasTyc 072133 in the H$\alpha$ region (thick lines). The inactive 
templates built up with rotationally broadened E{\sc lodie} Archive spectra are displayed with thin lines. The two components of 
HD 183957 show only a small filling of their H$\alpha$ profiles, while RasTyc 072133
displays a double-peaked  H$\alpha$ emission profile.}
\label{fig:Halpha2}
\end{figure}

The two components of HD 183957 show only a small filling of their H$\alpha$ profiles (Fig.~\ref{fig:Halpha2}), 
while the other RasTyc stars display H$\alpha$ emission profiles with a variety of shapes, going from a simple 
symmetric emission profile (102623) to a double-peaked strong emission line (215940). It has been also observed
a very broad, complex feature with a filled-in core and an emission blue wing (072133). 
A H$\alpha$ profile similar to that displayed by the latter star has been sometimes 
observed in some long-period RS CVn's, like HK Lac (e.g. Catalano \& Frasca 1994). RasTyc 072133 was 
classified as a semi-regular variable after Hipparcos, but it displays all the characteristics of a 
RS~CVn SB1 binary.
The E{\sc lodie} spectra of 221428 in the H$\alpha$ region show that the secondary (less massive) 
component displays a H$\alpha$ line always in emission with a stronger intensity around phase 0$\fp$7.
The  OAC spectra of DF~Cam always display a pure H$\alpha$ emission line, whose intensity  varies with the 
orbital/rotational phase. Similarly to 072133, DF~Cam, considered as a semi-regular variable after Hipparcos 
photometry, is very likely an active binary of the RS~CVn or BY~Dra class.

The equivalent width of the lithium $\lambda$6708 line, $EW_{\rm Li}$, was measured on the 
E{\sc lodie} spectra.  For the three sources for which we were able to detect and measure 
$EW_{\rm Li}$, we deduced lithium abundance, $\log N(Li)$, in the range 1.3--1.8, 
according to \cite*{Pav96} NLTE calculations.

\acknowledgements

We are grateful to the members of the staff of OHP and OAC observatories for their support and help
with the observations.
This research has made use of SIMBAD and VIZIER databases, operated at CDS, 
Strasbourg, France.


\begin{thebibliography}{}

\bibitem[\protect\astroncite{Baranne et al.}{1996}]{Bar96}  
Baranne A., Queloz D., Mayor M., et al., 1996, A\&AS 119, 373
\bibitem[\protect\astroncite{Catalano et al.}{1994}]{Catalano94} 
Catalano S. and Frasca A. 1994, A\&A 287, 575
\bibitem[\protect\astroncite{Frasca et al.}{2003}]{Frasca03}
Frasca A., Alcal\`a J.M., Covino E., Catalano S., Marilli E. and Paladino R. 2003, A\&A 405, 149
\bibitem[\protect\astroncite{Guillout et al.}{1999}]{Guillout99}
 Guillout P., Schmitt J. H. M. M., Egret D., Voges W., Motch C. and Sterzik M. F. 1999, A\&A 351, 1003
\bibitem[\protect\astroncite{Katz et al.}{1998}]{Katz98}
 Katz D., Soubiran C., Cairel R., Adda M. and Cautain R. 1998, A\&A 338, 151
\bibitem[\protect\astroncite{Landolt}{1992}]{Lan92}       
Landolt, A. U. 1992, AJ, 104, 340
\bibitem[\protect\astroncite{Lo Presti \& Marilli}{1993}]{LoPr93}      
Lo Presti, C., \& Marilli, E. 1993,  PHOT. Photometrical Data Reduction Package. 
Internal report of Catania Astrophysical Observatory N.~2/1993 			    
\bibitem[\protect\astroncite{Lucy \& Sweeney}{1971}]{Lucy71}
Lucy, L. B. and Sweeney, M. A., 1971, AJ 76, 544
\bibitem[\protect\astroncite{Pavlenko \& Magazz\`u}{1996}]{Pav96}
Pavlenko Y.V. \& Magazz\`u A. 1996, A\&A 311, 961
\bibitem[\protect\astroncite{Prugniel \& Soubiran}{2001}]{Prugniel01}
Prugniel, P. and Soubiran, C. 2001, A\&A 369, 1048
\bibitem[\protect\astroncite{Queloz et al.}{1998}]{Queloz98}
 Queloz D., Allain S., Mermilliod J.-C., Bouvier J. and  Mayor, M. 1998, A\&A 335, 183 


\end{thebibliography}
\end{document}